\documentclass[aps,pre,twocolumn,showpacs,notitlepage,floatfix,superscriptaddress]{revtex4-1}

\usepackage{amsmath}
\usepackage{amssymb}
\usepackage{mathtools}
\usepackage[all]{xy}
\input xy
\xyoption{matrix}
\xyoption{2cell}
\xyoption{arrow}
\xyoption{curve}
\xyoption{all}

\usepackage{graphicx}
\usepackage{epsfig}
\usepackage{dcolumn}
\usepackage{bm}
\usepackage{amsmath}
\usepackage{graphicx}
\usepackage[latin1]{inputenc}
\usepackage{ulem}
\usepackage{epstopdf}
\usepackage{subfigure}
\usepackage{color}

\RequirePackage[usenames,dvipsnames]{xcolor}
\usepackage{soul}

\newcolumntype{P}[1]{>{\centering\arraybackslash}p{#1}}

\begin{document}
\title{
Intermittent many-body dynamics at equilibrium
}

\author{C. Danieli}
\affiliation{New Zealand Institute for Advanced Study, Centre for Theoretical Chemistry \& Physics, Massey University, Auckland, New Zealand}
\affiliation{Center for Theoretical Physics of Complex Systems, Institute for Basic Science, Daejeon, Korea}
\author{D.K. Campbell}
\affiliation{Boston University, Department of Physics, Boston, Massachusetts 02215, USA}
\author{S. Flach}
\affiliation{Center for Theoretical Physics of Complex Systems, Institute for Basic Science, Daejeon, Korea}
\affiliation{New Zealand Institute for Advanced Study, Centre for Theoretical Chemistry \& Physics, Massey University, Auckland, New Zealand}


\date{\today}

\begin{abstract}

The equilibrium value of an observable defines a manifold in the phase space of an ergodic
and equipartitioned many-body system. A typical trajectory pierces that manifold infinitely often as time goes to infinity. We use these piercings to measure both the relaxation time of the lowest frequency eigenmode of the Fermi-Pasta-Ulam chain (FPU), as well as the fluctuations of the subsequent dynamics in equilibrium.
The dynamics in equilibrium is characterized by a power-law distribution of excursion times far off equilibrium, with diverging variance. Long excursions arise from sticky dynamics close to q-breathers localized in normal mode space. 
Measuring the exponent allows to predict the transition into nonergodic dynamics. We generalize our method to Klein-Gordon lattices (KG) where the sticky dynamics is due to discrete breathers localized
in real space.

\end{abstract}

\maketitle

Equipartition and thermalization have been central research topics in many-body interacting systems since the time of Maxwell, Boltzmann and Gibbs. The first computer experiment, aimed to observe equipartition starting from a microscopic reversible dynamical system, was carried out in the 1950s by Enrico Fermi, John Pasta, Stanislaw Ulam
and Mary Tsingou \cite{Fermi55}. Now famous as the Fermi-Pasta-Ulam (FPU) paradox (for reviews see \cite{Ford,Weissert,Galavotti,Porter2009}), this experiment failed to find equipartition but instead revealed intriguing nonlinear dynamics - including the celebrated {\it FPU recurrences} \cite{Fermi55} - which has challenged and puzzled researchers for more than 60 years (for a recent survey of the state of the art, see \cite{Galavotti}). In brief, attempts to understand the full dynamics, including the recurrences, led to the observation (and naming) of {\it solitons} \cite{zabuskykruskal, zabuskydeem} and important developments in Hamiltonian chaos \cite{izrailevchirikov}.
It is now known that these unexpected recurrences are linked to the choice of initial conditions used by FPU, which are set close to exact coherent time-periodic (or even quasiperiodic) trajectories, 
{\it e.g.} {\it q-breathers}, which show exponential localization of energy in {\it normal mode space} \cite{Flach06, Christodoulidi10}. 
Even if these trajectories  have support of measure zero in the phase space, they might have a finite measure impact simply by being linearly stable \cite{Flach06}.
Several other studies admit coherent time-periodic states localized in real space, which are known as {\it discrete breathers} or {\it intrinsic localized modes} \cite{discretebreathers} 
 and exist e.g. in Klein-Gordon (KG) lattices \cite{ivanchenko04}. 
These states can also be linearly stable and thus may have finite measure impact.
Importantly, both discrete breathers and q-breathers have been experimentally observed in a large variety of physical settings \cite{discretebreathers,qbexp}.
Thus the central question becomes: How does the presence of such coherent states of measure zero affect the dynamical properties of a {\it thermalized} many-body system?  How do they affect the possible transition from 
ergodic to a non-ergodic dynamics?
Interestingly there are only a few recorded numerical attempts 
to address this complex issue \cite{escande94,tsironis96,rasmussen00,eleftheriou,matsuyama15,zhang16,mulhern15,gershgorin05}.  
In our view, this is the result of the lack of a clear strategy which can go beyond the analysis of correlation functions
( which obscure the understanding of a detailed correspondence between the equilibrium dynamics and coherent structures due to event averaging).

Given a many-body system which possesses linearly stable coherent states, we choose an observable $f$ ({\it i.e.}, some
function of the phase space variables) whose value is sensitive to the excitation of such states. We 
assume that the many-body system is thermalizing,
or ergodic, {\it i.e.} that the phase space trajectory is evolving under the constraint of fixed total energy (and perhaps other conserved quantities)
such that the time average $\langle f \rangle_t \equiv \langle f \rangle$ is independent of the actual chosen trajectory, up to a set of measure zero (like periodic orbits, which can
persist even in the strongest chaotic flows). The actual value of $f(t)$ will depend upon time $t$ along a typical trajectory. As time goes to infinity the trajectory is then forced to 
pierce infinitely often
a submanifold $\mathcal{F}_f$ of codimension 1 which hosts all phase space points with $f \equiv \langle f \rangle $. 
The submanifold can be considered as a generalized ergodic Poincar\'e section, which is fixed by the choice of $f$, the integrals of motion and the assumption of ergodicity.
The time intervals between
consecutive piercings will carry the information on whether (and when) the trajectory was visiting a sticky region in phase space. Hence we will study the statistics of these time intervals.
In contrast to a correlation function, these are the statistics of trackable events and will always permit us 
to return to the event of interest,
in order to inspect it microscopically. With 
this insight we also arrive at a novel quantitative dynamical characterization of the degree of equipartition of a given microscopical state, {\it i.e.} a point on the considered trajectory. Rather than using an entropy-like measure ({\it e.g.} the distance from the set $\mathcal{F}_f$),
it is the time the trajectory needs to reach and pierce $\mathcal{F}_f$ which will decide whether the given configuration is close to or far from equilibrium.

We apply the above ideas to both FPU and KG systems with the Hamiltonian function of the canonically conjugated pairs of real space momenta
and coordinates  $\{p_n,q_n\}$
\begin{align}
\label{eq:FPU}
H = \sum_{n=0}^{N}&\left[ \frac{p_n^2}{2} + V(q_n) + W(q_{n+1} - q_n)\right]\ , \\
\label{eq:FPU_2}
\text{FPU:}\quad &V(q)=0\ ,\quad W(q)= \frac{1}{2}q^2 + \frac{\alpha}{3}q^3 \ , \\ 
\label{eq:KG_2}
\text{KG:}\quad \  &V(q)=\frac{1}{2}q^2 + \frac{1}{4}q^4\ ,\quad W(q)= \frac{k}{2}q^2 \ .
\end{align}  
Both models turn into integrable sets of noninteracting normal modes in the limit of vanishing energies. In addition the KG system turns into an integrable set of noninteracting anharmonic oscillators in the limit of diverging energies, due to its onsite anharmonicity (as opposed to the FPU case).
We use fixed boundary conditions $p_0=p_{N+1}=q_0=q_{N+1}=0$ for the FPU chain in line with Ref. \cite{Fermi55},.  For the KG chain we use instead periodic boundary conditions $p_1=p_{N+1}$, $q_1=q_{N+1}$ in order to
keep all sites equivalent and to avoid edge effects.

To address the normal mode dynamics, we  use the FPU system and the canonical transformation
\begin{equation}
\begin{split}
\left(
\begin{array}{c}
P_k\\
Q_k\\
\end{array}
\right)
&= \sqrt{\frac{2}{N+1}}\sum_{n=1}^{N}
\left(
\begin{array}{c}
p_n\\
q_n\\
\end{array}
\right)
 \sin \bigg(\frac{\pi n k}{N+1}\bigg)
\end{split}
\label{eq:Fourier coordinates}
\end{equation}
with $k=1,\dots, N$, which
diagonalizes the harmonic part of $H$ ($\alpha=0$ in (\ref{eq:FPU_2})) with the normal mode momenta and coordinates $\{P_k,Q_k\}$. The mode energies and frequencies are
\begin{equation}
E_k = \frac{P_k^2 + \omega_k^2 Q_k^2}{2}\ , \quad \omega_k = 2\sin\left(\frac{\pi k }{2(N+1)}\right)\;.
\label{eq:modes_energy}
\end{equation}
For $\alpha\neq 0$ the mode energies become time-dependent 
and are monitored using
the normalized distribution $\nu_k(t) = E_k(t) / \sum_{k=1}^{N}E_k(t)$, with $\sum_k \nu_k=1$.
A common tool to monitor the degree of inhomogeneity of the distribution is the spectral entropy \cite{Casetti97,onorato15}
\begin{equation}
S(t) = - \sum_{k=1}^{N} \nu_k(t) \ln(\nu_k(t))
\label{eq:spectral entropy}
\end{equation}
with $0\leq S \leq S_{max}=\ln N$. Its rescaled analogue is
\begin{equation}
\eta(t) = \frac{S(t) - S_{max}}{S(0) -  S_{max}} \; , \; 0 \leq \eta \leq 1 \;.
\label{eq:rescaled entropy}
\end{equation}

To address the real space dynamics of the KG system, we use the energy densities
\begin{equation}
\epsilon_n = \frac{p_n^2}{2} + V(q_n) + \frac{k}{4}\sum_{s=\pm1} W(q_{n+s} - q_n)\ .
\label{KG_Edensities}
\end{equation}
An equally common measure of energy distribution inhomogeneity is the participation number $P$, which yields the number of strongly excited renormalized energies $\mu_n(t) = \epsilon_n(t) / \sum_{n=1}^{N}\epsilon_n(t)$:
\begin{equation}
P^{-1}(t) = \sum_{n=1}^N \mu_n^2(t) \;,\;1 \leq P \leq N\;.
\label{eq:Pn}
\end{equation}
\begin{figure}[!h]
\centering
\includegraphics[ width=1.025\columnwidth]{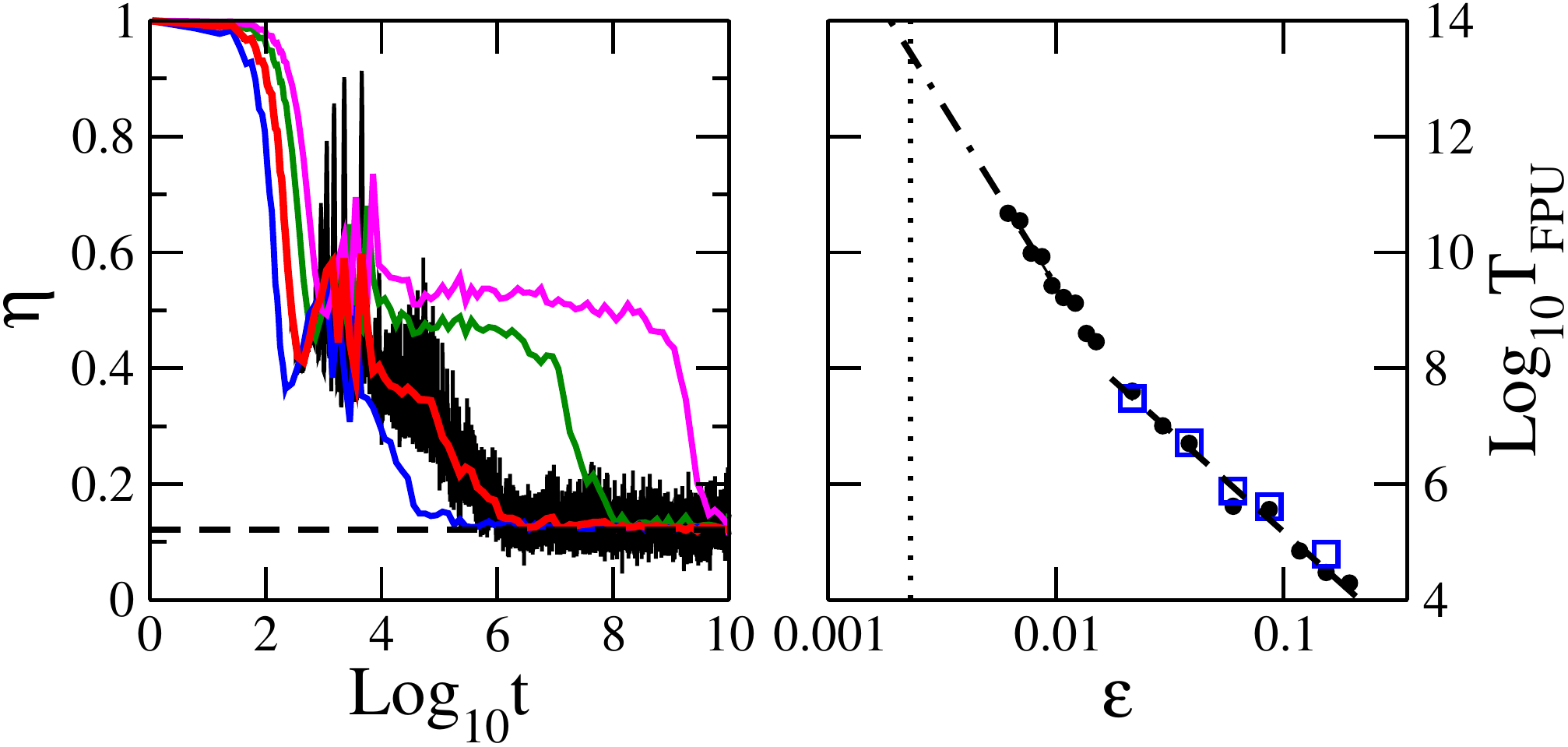}
\caption{Left: instantaneous (black) and window-averaged (red) time evolution of the entropy $\eta$ for $\epsilon =0.0566$, and window-averaged time evolution for $\epsilon=0.145$ (blue), $\epsilon=0.0566$ (red), $\epsilon=0.0204$ (green) and $\epsilon=0.0091$ (magenta). Black dashed line:  $\langle\eta\rangle=0.1218$. 
Right: $T_{\text{FPU}}$ (black circles) vs. $\epsilon$. The blue squares are the data from Ref. \cite{Casetti97}. The black dashed and dashed-dotted lines guide the eye and indicate a crossover at $\epsilon \approx 0.01$. Vertical dotted line: $\epsilon=0.0023$.}
\label{fig1}
\end{figure} 
Both observables $\eta$ and $P^{-1}$ will fluctuate along the temporal evolution of a trajectory. Let us assume that their averages $\langle\eta\rangle$, $\langle P^{-1}\rangle$ exist and can be computed using the Gibbs distribution (which follows from well-known general considerations of counting microstates or maximizing the entropy)
\begin{equation}
W_B = \frac{1}{Z} {\rm e}^{-\beta H}\;,\; Z = \int_\Gamma {\rm e}^{-\beta H} d\Gamma \; .
\label{eq:Boltzman_distro}
\end{equation}
Here $\Gamma$ denotes the whole available phase space, and $\beta$ is the inverse temperature.
At low enough energies the anharmonic energy contribution for the FPU system 
will be a small correction and can be neglected when computing averages; 
its relevance is reduced to the highly important nonlinear mode interaction which is the crucial source of deterministic chaos and equipartition.
The final integration using the Gibbs distribution (\ref{eq:Boltzman_distro}) can be performed analytically \cite{Goedde}:
\begin{equation}
\langle \eta \rangle = \frac{1-\gamma}{\ln N - S(0)} \;, 
\end{equation}
where $\gamma\approx0.5772$ is the Euler constant.
For the KG system, we obtain the average $\langle P^{-1}\rangle$ directly by numerically averaging until the total integration time $T=10^8$. 
These averages define the equilibrium manifolds $\mathcal{F}_{\eta},\mathcal{F}_P$
which we will use for the subsequent analysis. 

The original FPU computation \cite{Fermi55} was performed for $N=32$ particles with only the lowest frequency mode excited, $Q_1\neq 0$ only. Then $S(0)=0$
and $\langle\eta\rangle\approx0.1218$.
We will benchmark our data with the results from \cite{Casetti97}, who used an {\it ad-hoc} value $\eta=0.1$.
The trajectory starts with $\eta (0) = 1 \gg \langle \eta \rangle$
close to a regular periodic orbit localized in momentum space (a $q$-breather) \cite{Flach06}.
A central target of many FPU paradox studies was to quantify the time this initial state needs to reach equipartition, if it ever does ({\it e.g.} \cite{Fermi55,Casetti97,Ponno11,onorato15, Deluca99}). Since equipartition means equal mode energies on average, we define the FPU equipartition time $T_{\text{FPU}}$
as the time the trajectory needs to reach the corresponding manifold $\mathcal{F}_{\eta}$.
We continue our computations beyond this equipartition time.
The trajectory has to cross the manifold $\mathcal{F}_{\eta}$ infinitely often, and we record the piercing times $t_i$ with $i\geq 1$ (note that $T_{\text{FPU}}\equiv t_1$).
The {\it return times} 
\begin{equation}
t_r(i) = t_{i+1} - t_{i}\ ,\quad i\geq 1\ 
\label{eq:rec_time_def}
\end{equation}
measure the time intervals the trajectory spends off the equilibrium manifold before piercing it again, with even and odd integers $i$ discriminating between corresponding excursions
into the two different phase space subspaces ({\it e.g.} $\eta>\langle\eta\rangle$ and $\eta<\langle\eta\rangle$).

The computations  
were carried out using a symplectic $SABA_2 C$ integrator with corrector \cite{Laskar01,Skokos09}, with a time step $\tau=0.1$; these choices keep the relative energy error of the order $10^{-5}$ \cite{supp}. 
The system size is $N=32$, and $\alpha=0.25$ in Eq.(\ref{eq:FPU_2}), and initial condition $P_k(0)=0$, $Q_k(0)=A\delta_{k,1}$, which translates into a corresponding 
total energy $E$, and energy density $\epsilon=E/N$ (see \cite{supp} for details). We follow the time dependence of observables and also perform a window averaging over a time window which is 100 times shorter than the actual running time.

In Fig.\ref{fig1} - left plot - we show the time evolution of the entropy $\eta$ for different energy densities $\epsilon$.
The curves start at the unity at $t=0$ (see Eq.(\ref{eq:rescaled entropy}))
and then settle to fluctuating intermediate values for a transient interval of time that increases
as the energy density $\epsilon$ decreases. Finally, at $t=T_{\text{FPU}}$ the observable
transits into fluctuations around equilibrium at values that approximate the Gibbs average $\langle\eta\rangle$ very well. 
The intermediate plateau corresponds to a {\it metastable} state, where all the mode energies $E_k$ are non-zero but assume an exponentially decaying profile \cite{Ponno11, Benettin11,Galavotti, Fucito82}. The second plateau corresponds to the regime of equipartition, confirming the validity of the Gibbs distribution.

In Fig.\ref{fig1} - right plot - we plot the FPU equipartition time $T_{\text{FPU}}$ as a function of the density $\epsilon$, along with the data from Ref.\cite{Casetti97}, which show very good agreement. We also satisfactorily compared our data to the extrapolated equipartition times from Ponno et al \cite{Ponno11}, see details in \cite{supp}. 
As noted previously, the equipartition time increases with decreasing energy density. Casetti {\it et al.} predicted the equipartition time at the original FPU energy density choice of $\epsilon=0.00226$ to be of the order of $T_{\text{FPU}}\approx10^{12}$ which currently 
requires about 30 days of CPU time with our system \cite{supp}.
However, the equipartition time shows a crossover at $\epsilon\approx0.01$. which was not reached by previous computations.
A straight-forward extrapolation from this crossover (see dashed-dotted line in In Fig.\ref{fig1} - left plot) increases this time to $T_{\text{FPU}}\approx10^{14}$ or about 10 years of CPU time on our system. 
Remarkably
the answer to whether the original FPU trajectory is thermalizing or not remains a very hard computational problem more than six decades 
after the first observation of the FPU paradox.

\begin{figure}
 \centering
 \includegraphics[ width=0.9\columnwidth]{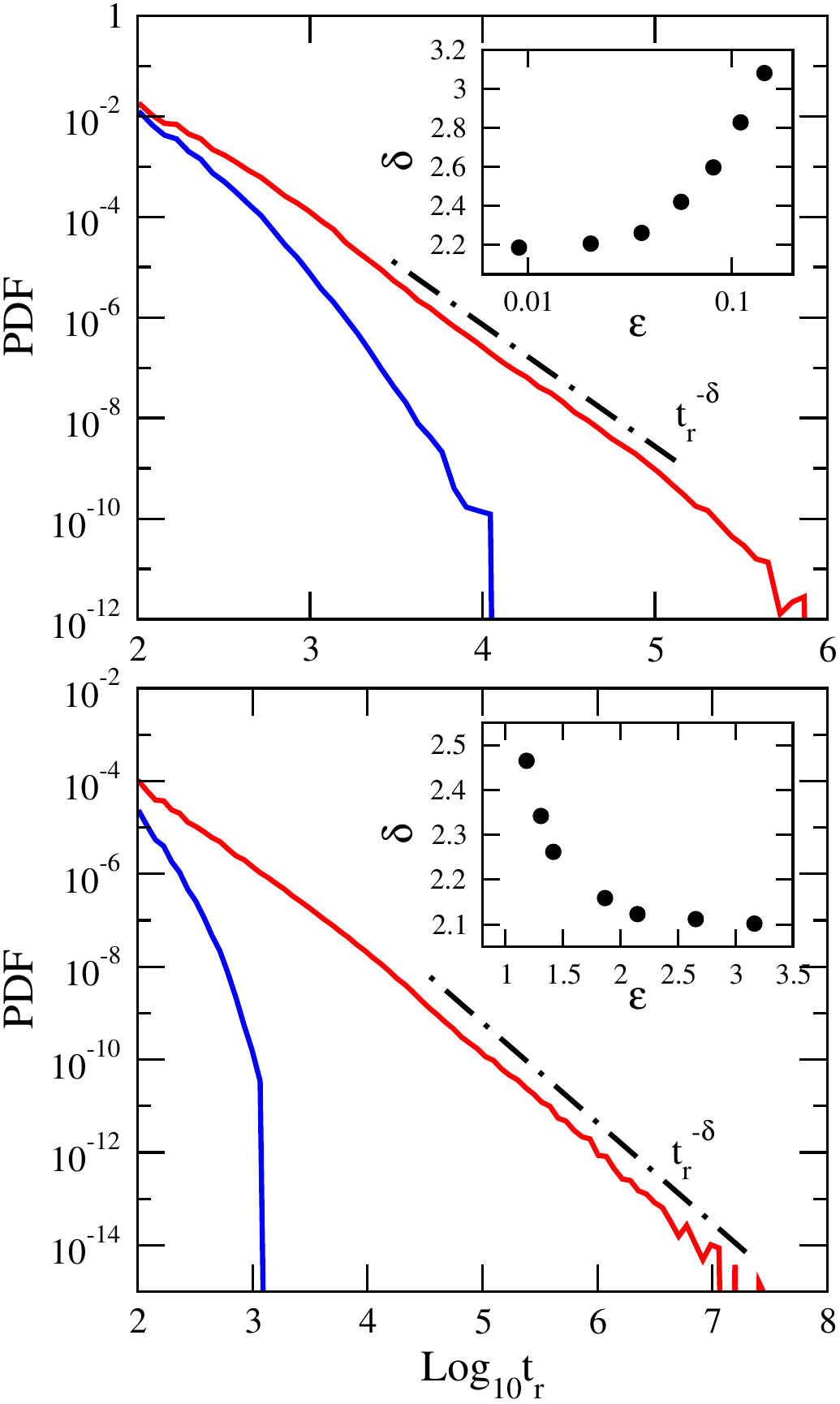}
 \caption{
PDFs $\mathcal{P}_{\pm}(t_r)$ for $\epsilon = 0.0566$ (FPU - upper plot) and $\epsilon=1.867$ (KG - lower plot). 
For both FPU and KG, the red (upper) curve corresponds to $\mathcal{P}_{+}(t_r)$ and the blue (bottom) one to $\mathcal{P}_{-}(t_r)$.
The dashed-dotted lines guide the eye and indicate the algebraic tails.
Inset: the exponent $\delta$ of the algebraic tails versus the energy density $\epsilon$.}
  \label{fig4}
\end{figure}

Let us now turn
to the analysis of the equilibrium dynamics beyond the equipartition time. We compute the sets of 
return times (\ref{eq:rec_time_def}) separately for the two different subspaces $\eta>\langle \eta \rangle$ and $\eta<\langle \eta \rangle$.
The probability distribution functions (PDFs) of these sets $\mathcal{P}_{\pm}(t_r)$ are shown for $\epsilon=0.0566$ in Fig.\ref{fig4} - upper plot.
In the subspace $\eta > \langle \eta \rangle$, the dynamics exhibits 
algebraic tails in the PDF $\mathcal{P}_+(t_r) \sim t_r^{-\delta}$ with an exponent $2 < \delta < 3$, which indicates a finite average (1st moment) $\langle t_r\rangle$ but a diverging variance (2nd moment) $\langle t_r^2\rangle$ (see \cite{supp} on the numerical details of estimating $\delta$).
The exponent $\delta$ decreases with decreasing energy density $\epsilon$, signalling the reaching of the integrable 
harmonic oscillator chain limit.
Note that for $\delta \leq 2$ the average $\langle t_r\rangle$ would diverge, and the ergodicity assumption would be 
violated, again indiciating the transition into a nonergodic, perhaps integrable, case.
Therefore, our method is sensitively predicting the transition from ergodic to nonergodic dynamics.
In contrast,
the subspace $\eta < \langle \eta \rangle$ dynamics yields tails in $\mathcal{P}_-(t_r)$ with finite moments; the tails are faster than algebraic but slower than exponential, presumably 
exponentials dressed with a power law. This is due to that subspace hosting microstates for which the normal modes are even more equipartitioned than on a Gibbs average. Such microstates have small probability, and are insensitive
for detecting nonequilibrium fluctuations.

We extend the analysis of the dynamics at equipartition and the distribution of the return times to the KG chain, a model known to posses discrete breather solution in the real space \cite{ivanchenko04}, and should show a related  transition to nonergodicity and integrability with increaasing energy density.
At variance to the FPU case, we will
search for a gradual loss of ergodicity upon {\sl increasing} the energy density, which should favour the excitations of 
discrete breathers.
We choose $N=32$, $k=0.1$, periodic boundary conditions, and random initial conditions with a predefined energy density.
We compute the time evolution of the participation ratio $P$ until total integration time $T=10^{10}$, and we record the return times $t_r$ between two consecutive piercings of the equilibrium manifold $\mathcal{F}_P$ again separating the  phase space in $P^{-1}> \langle P^{-1} \rangle$ and $P^{-1}< \langle P^{-1} \rangle$.
The PDFs $\mathcal{P}_{\pm}(t_r)$ obtained for energy density $\epsilon=1.867$ are shown in Fig.\ref{fig4} - lower plot - where, for $P^{-1}> \langle P^{-1} \rangle$, the algebraic tail of $\mathcal{P}_{+}(t_r)  \sim t_r^{-\delta}$ is visible while $\mathcal{P}_{-}(t_r)$ shows exponential cut-off.
In the inset we plot the exponent $\delta$, which drops below
values of $\delta=3$ and continues to decrease towards the critical case $\delta = 2$ with increasing energy density
$\epsilon$. Similar to the FPU case, the KG system dynamics shows a transition from ergodic into nonergodic dynamics.

\begin{figure}[h]
 \centering
 \includegraphics[ width=0.525\columnwidth]{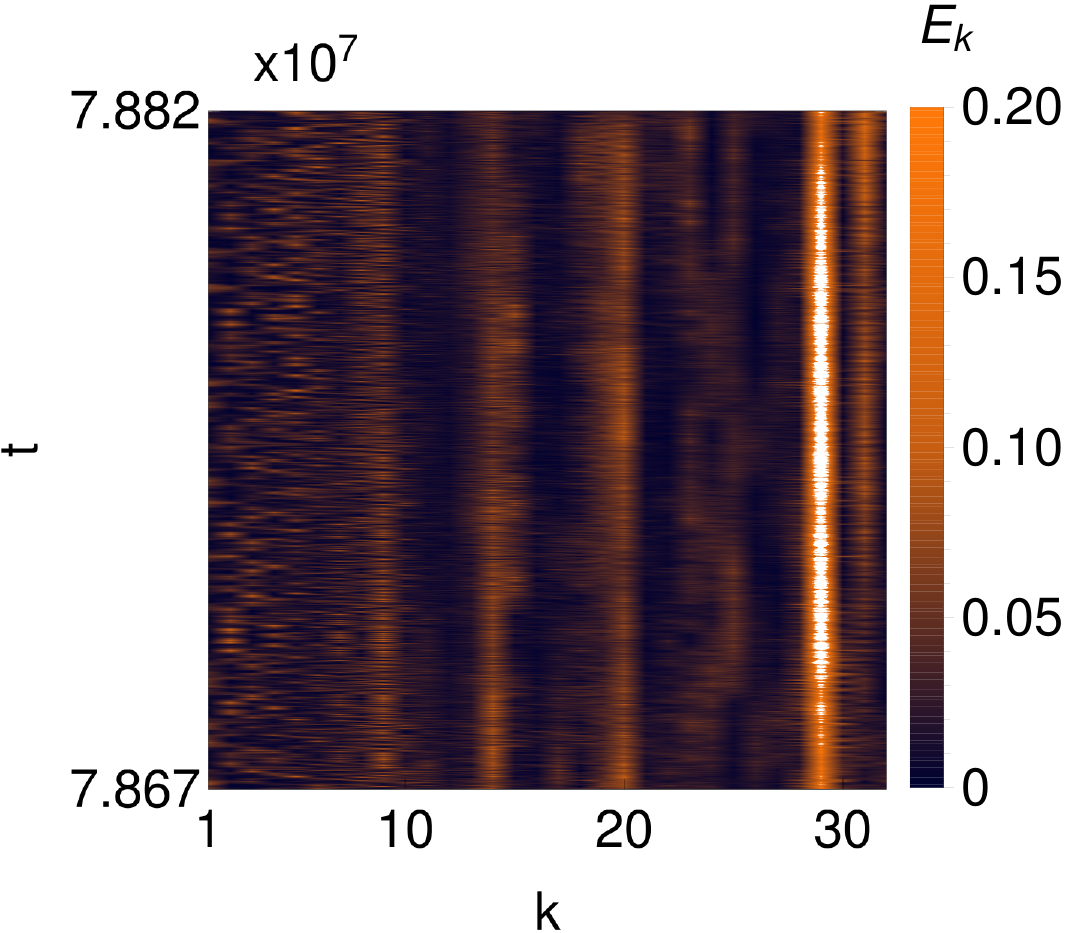}
\hspace{-1.8mm} 
 \includegraphics[ width=0.47\columnwidth]{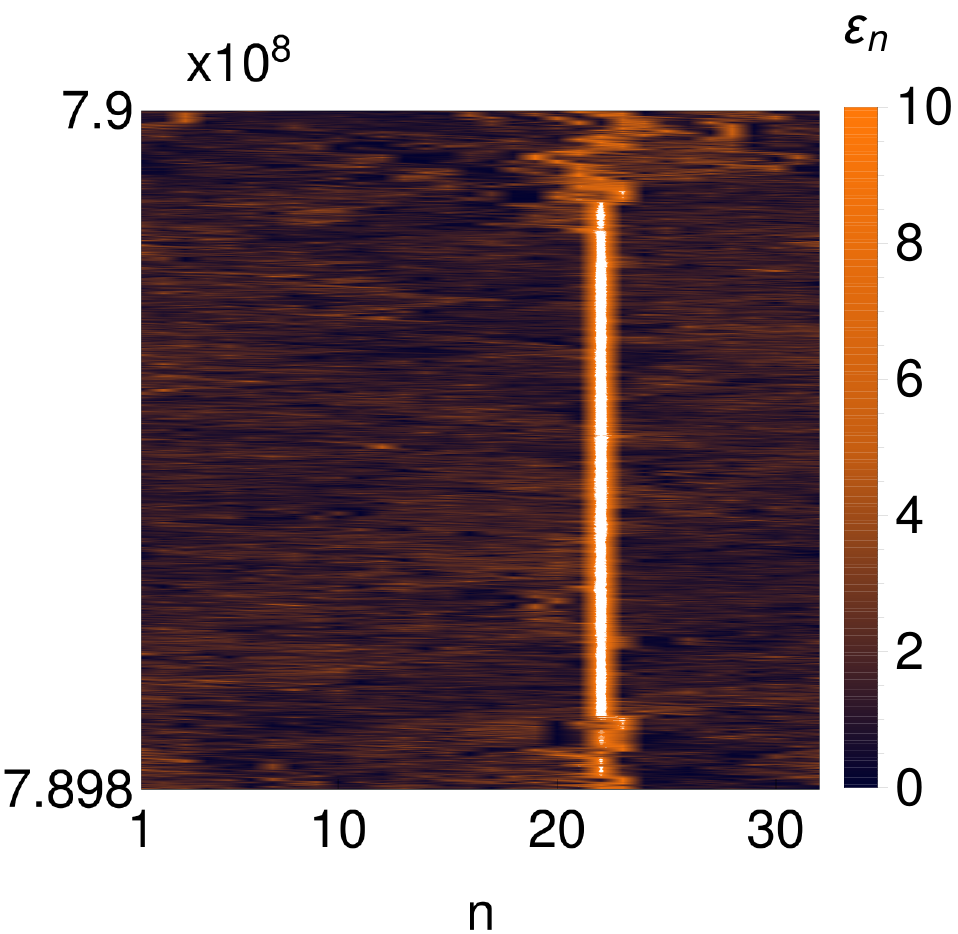}
 \caption{
Left: mode energies $E_k$ of the FPU as function of time during one of the longest trapping events for 
$\epsilon = 0.0566$. Right: 
energy densities $\epsilon_n$
 of the KG as function of time during one of the longest trapping events for 
$\epsilon = 1.748$.
}
 \label{fig3}
\end{figure}

The algebraic tails of the PDF of the 
return times with $3 > \delta > 2$ imply that the trajectory is with high probability getting trapped in some parts of phase space 
for long times,
whose average is finite, but whose variance diverges. We conjecture that these trapping events are due to visiting regions of the phase space which are substantially close to some regular orbits.
In order to substantiate this conjecture, we show in Fig.\ref{fig3} the time evolution of the mode energies $E_k$ (FPU - left plot) and the energy densities $\epsilon_n$ (KG - right plot) during one of their longest excursions far from equilibrium. 
At the beginning of the event we observe the focusing of energy in one of the modes (FPU) or sites (KG) respectively.
These breather-like excitations then survive over the entire duration of the excursion, only to dissolve their energy back
into the other degrees of freedom at the end of the event
(for further details, see \cite{supp}).

To further substantiate our observation, we show the correlation between the $1^{st}$ moment of the event-averaged mode energy distribution $C = \sum_{k=1}^{N}k \nu_k$ for the FPU case
versus the trapping event time $t_r$. 
\begin{figure}
 \centering
 \includegraphics[ width=\columnwidth]{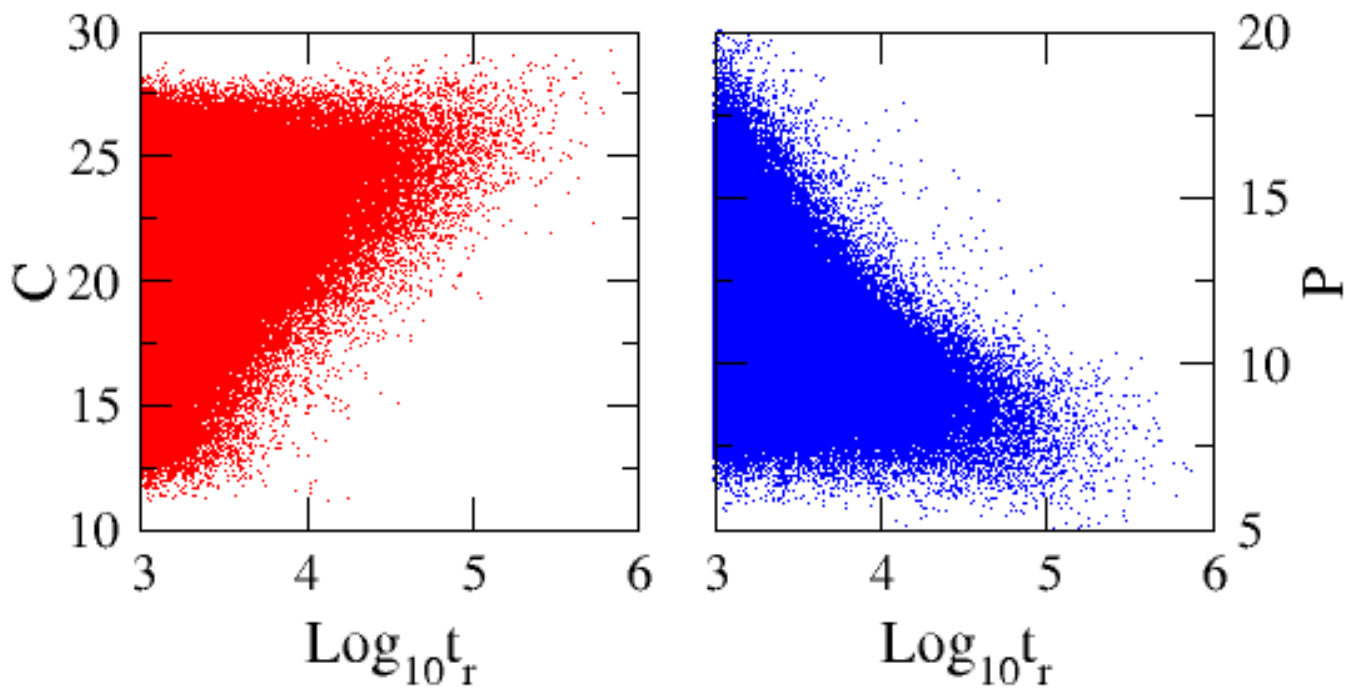}
 \caption{
$C(t_r)$ (left plot) and $P(t_r)$(right plot) for $\epsilon=0.0566$ (see text for details).  
The broad scattering of data is due to many independent events yielding similar return times $t_r$.
}
 \label{fig5}
\end{figure}
In Fig.\ref{fig5} (left plot), we observe that large return times $t_r$ imply large values of $C \approx N$, signalling a tendency towards
high frequency excitations. 
Most importantly the corresponding computation of the participation number $P$ of the
event-averaged mode energy distributions in Fig.\ref{fig5} (right plot) shows that large return times 
correlate with smaller values of $P$, a typical
case of a strongly inhomogeneous distributions.
Therefore the equilibrium FPU dynamics produces sticky excursions with long duration to strongly excited high frequency modes.

The properties of fluctuations in equilibrium should not depend on the choice of the trajectory, in accord with the assumption of equipartition and ergodicity. We tested that in the FPU chain by launching various other trajectories, e.g. exciting one high frequency mode, or several modes with different frequencies (not shown here). We observed that the statistics of return times is universal and not depending on the choice of the initial state.

Algebraic tails in correlation functions or distributions of trapping times have been previously studied for low-dimensional
dynamical systems with a mixed phase space \cite{mixed1,mixed2}, and related to the hierarchic fractal structure
of the phase space at regular island boundaries, similar to the phenomenological approach to understand glassy dynamics. However higher phase space dimensions destroy the simple mixed phase space picture, preventing the use of this simple argument for the observation of algebraic tails \cite{higher}.
In the present work we derive a well-defined sectioning at equilibrium, and a clear interpretation of the presence of
algebraic tails in terms of temporal excitation of coherent states, like time-periodic q-breathers.
The large phase space dimension does not easily allow to connect to the physics of glasses, since the potential functions
are smooth, and invariant regular trajectories. Instead we are in need of a new understanding how regular states
of measure zero (e.g. time-periodic solutions) can act as dynamical barriers and bottlenecks in high-dimensional
phase spaces.

We arrived at a general method to analyze the relaxation from non-equilibrium states and the equilibrium fluctuations of interacting many-body systems.
The essence is to identify the relevant coherent excitations which will be the cause of stickiness, and to choose a proper observable $f$ which can detect these events.
The corresponding equilibrium value $\langle f \rangle$ defines the co-dimension 1 equilibrium manifolds, and the subsequent statistical analysis of the distributions of equilibrium fluctuations. 
When algebraic tails are observed in contrast to exponential cutoffs, the divergence of suitably high moments of the distribution indicates sticky dynamics. When the exponent $\delta < 3$, the non-equilibrium excursions into
sticky events start to dominate the dynamics. Finally when $\delta \leq 2$ the first moment diverges indicating the
loss of ergodicity altogether.
We expect therefore that our method can be used for a broad set of other cases where nonergodic fluctuations affect the dynamics of many-body systems , such as
ultracold atomic gases in optical potentials approximated by the discrete Gross-Pitaevsky equation, or
networks of weakly interacting superconducting grains, among others.
\begin{acknowledgements} 
We thank Peter Jeszinszki and Ihor Vakulchyk for helpful discussions on computational aspects.
The authors acknowledge financial support from IBS (Project Code:IBS-R024-D1).
\end{acknowledgements}



\widetext
\clearpage
\begin{center}
\textbf{SUPPLEMENTAL MATERIAL}
\end{center}

\section{The $SABA_2 C$ symplectic integratior} \label{sec:SABA}

To integrate the FPU and the KG chains in time, we use a symplectic integration scheme $SABA_2C$ proposed in \cite{Laskar01}, which is an improved version of the $SABA_2$ scheme. The $SABA_2$ scheme consists in separating the Hamiltonian $H$ in two integrable parts $H = A+B$, each one with solutions $e^{ t  L_{A}}$ and $e^{ t  L_{B}}$. The $SABA_2$ scheme of integrating the coupled differential equations of motion is defined as
\begin{equation}
SABA_2 = e^{ c_1\tau L_{A}} e^{ d_1\tau L_{B}} e^{ c_2\tau L_{A}} e^{ d_1\tau L_{B}} e^{ c_1\tau L_{A}}
\label{eq:saba2}
\end{equation} 
where $c_1= \frac{1}{2}\big(1-\frac{1}{\sqrt{3}}\big)$ and $c_2=\frac{1}{\sqrt{3}}$ while  $d_1=\frac{1}{2}$.  The corrector term $e^{\tau L_C}$ is the solution of the Hamiltonian $C=\{ \{ A,B\},B\}$, which adds two more matrix operations:
\begin{equation}
SABA_2 C = e^{ -(\tau^3 g/2) L_{C}} (SABA_2) e^{ -(\tau^3 g/2) L_{C}}
\label{eq:SABA2C}
\end{equation}
where $g=\frac{2-\sqrt{3}}{24}$. Further details can be found in \cite{Laskar01,Skokos09}. 
In the FPU case, the Hamiltonians $A$ and $B$ are respectively   
\begin{equation}
A = \sum_{n=0}^{N} \frac{p_n^2}{2}\ ;\qquad  B = \sum_{n=0}^{N}   \bigg[\frac{(q_{n+1} -   q_n)^2}{2}  +\alpha \frac{(q_{n+1} - q_n)^3}{3}\bigg] \ ;
\label{eq:FPU_A}
\end{equation}
while the Hamiltonian $C$ is
\begin{equation}
C = \sum_{n=0}^{N}\Big[ \big(2q_n - q_{n+1} - q_{n-1}  \big) \big(1+\alpha (q_{n+1} - q_{n-1})\big)  \Big]^2\ .
\label{eq:C}
\end{equation} 
The operators $e^{ \tau  L_{A}}$, $e^{ \tau  L_{B}}$ and $e^{\tau L_C}$ propagate the coordinates ${\bf x}(t) = ({\bf p}(t),{\bf q}(t))  = ({\bf p},{\bf q}) $ at time $t$ to the coordinates ${\bf x}(t+\tau) = ({\bf p}(t+\tau ),{\bf q}(t+\tau)) = ({\bf p}',{\bf q}')$ at time $t+\tau$:
\begin{equation}
\begin{split}
&e^{ \tau L_{A}}: \left\{
\begin{array}{rl}
   q_n' &= p_n \tau + q_n         \\
  p_n' &= p_n
\end{array} \right. \ ; \\
&e^{ \tau L_{B}}: \left\{
\begin{array}{rl}
   q_n' &=   q_n         \\
  p_n' &=  \big[  q_{n+1} +  q_{n-1}  -  2 q_n   \big]
          \big[  1+\alpha\big( q_{n+1} -   q_{n-1}  \big) \big]\tau +  p_n
\end{array} \right.\ ;
\end{split}
\label{eq:etA_atB}
\end{equation}
\begin{equation}
e^{ \tau L_{C}}: \left\{
\begin{array}{rl}
   q_n' &=  q_n         \\
     p_1' &=   2 \Big\{2 \big(  q_{2}  -  2 q_n   \big) \big[  1+\alpha q_{2} \big]^2 -  q_{1}  \big[  1 +\alpha  q_{1}  \big] \big[  1 + 2\alpha  q_{1}  \big)\big]  \\
         & - \big(  q_{3} +  q_{1}  -  2 q_{2}   \big)  \big[  1 +\alpha\big( q_{3} -   q_{1}  \big)  \big]\big[  1 - 2\alpha\big( q_{1} -   q_{2}  \big)  \big] \Big\}\tau + p_1 \\
  p_n' &=   2 \Big\{2 \big(  q_{n+1} +  q_{n-1}  -  2 q_n   \big) \big[  1+\alpha\big( q_{n+1} -   q_{n-1}  \big) \big]^2 \qquad \qquad\quad n=2,\dots,N-1 \\
         & - \big(  q_{n+2} +  q_{n}  -  2 q_{n+1}   \big)  \big[  1 +\alpha\big( q_{n+2} -   q_{n}  \big)  \big]  \big[  1 - 2\alpha\big( q_{n} -   q_{n+1}  \big)  \big]\\
          & - \big(  q_{n-2} +  q_{n}  -  2 q_{n-1}   \big)   \big[  1 +\alpha\big( q_{n} -   q_{n-2}  \big)  \big]  \big[  1 - 2\alpha\big( q_{n-1} -   q_{n}  \big)\big]  \Big\}\tau + p_n \\
       p_{N}' &=   2 \Big\{2 \big(  q_{N-1}  -  2 q_{N}   \big) \big[  1-\alpha   q_{N-1}   \big]^2   -   q_{N}   \big[  1 -\alpha   q_{N} \big]  \big[  1 - 2\alpha  q_{N}  \big]\\
          & - \big(  q_{N-2} +  q_{N}  -  2 q_{N-1}   \big)   \big[  1 +\alpha\big( q_{N} -   q_{N-2}  \big)  \big]  \big[  1 - 2\alpha\big( q_{N-1} -   q_{N
          }  \big)\big]  \Big\}\tau + p_N    
\end{array} \right. 
\label{eq:etC}
\end{equation}
A detailed presentation of the symplectic integration scheme $SABA_2C$ can be found in the appendix of \cite{Skokos09}. 
\begin{figure}[h!]
 \centering
 {\includegraphics[ width=0.45\columnwidth]{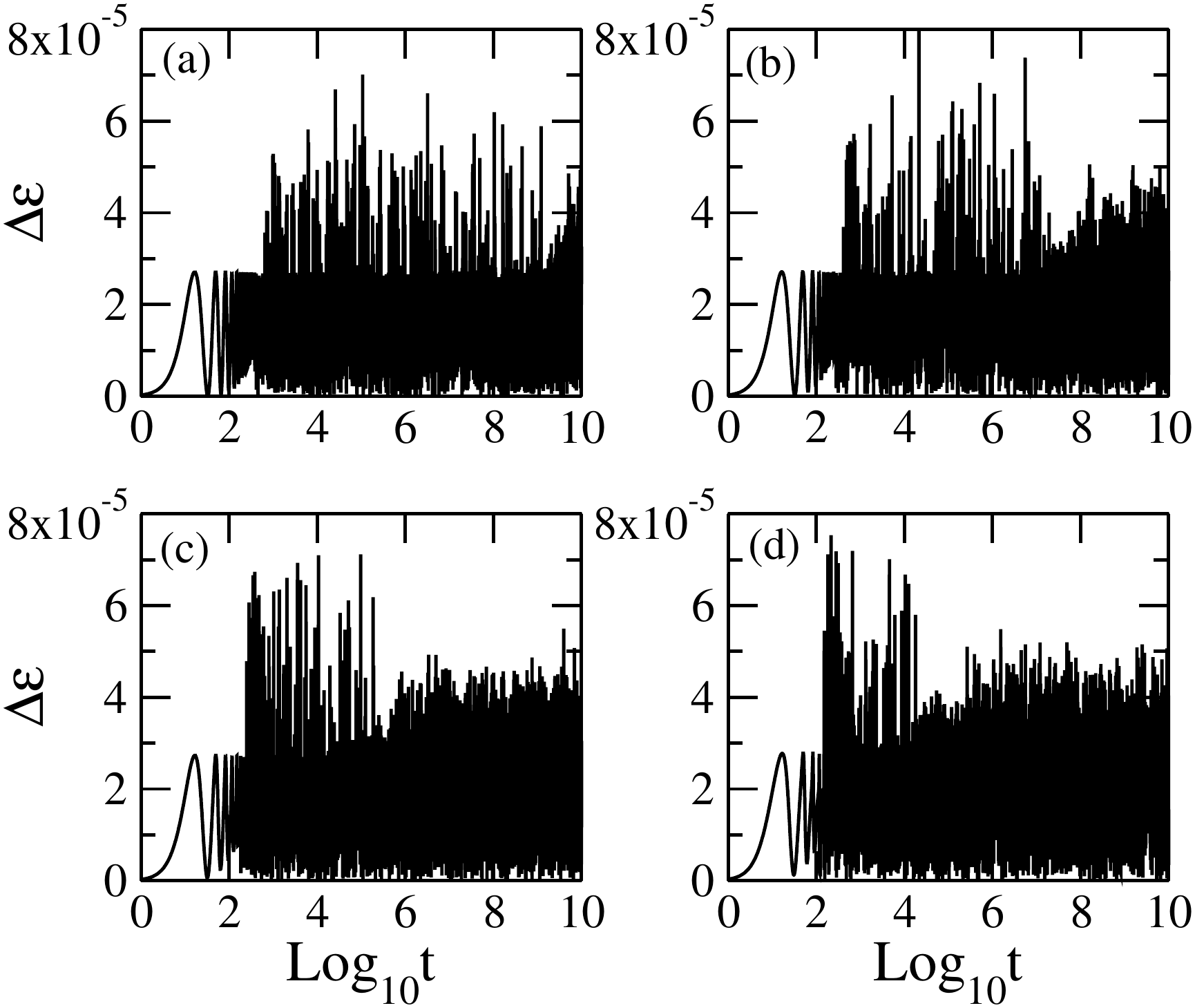}}
 \hspace{10mm}
 {\includegraphics[ width=0.45\columnwidth]{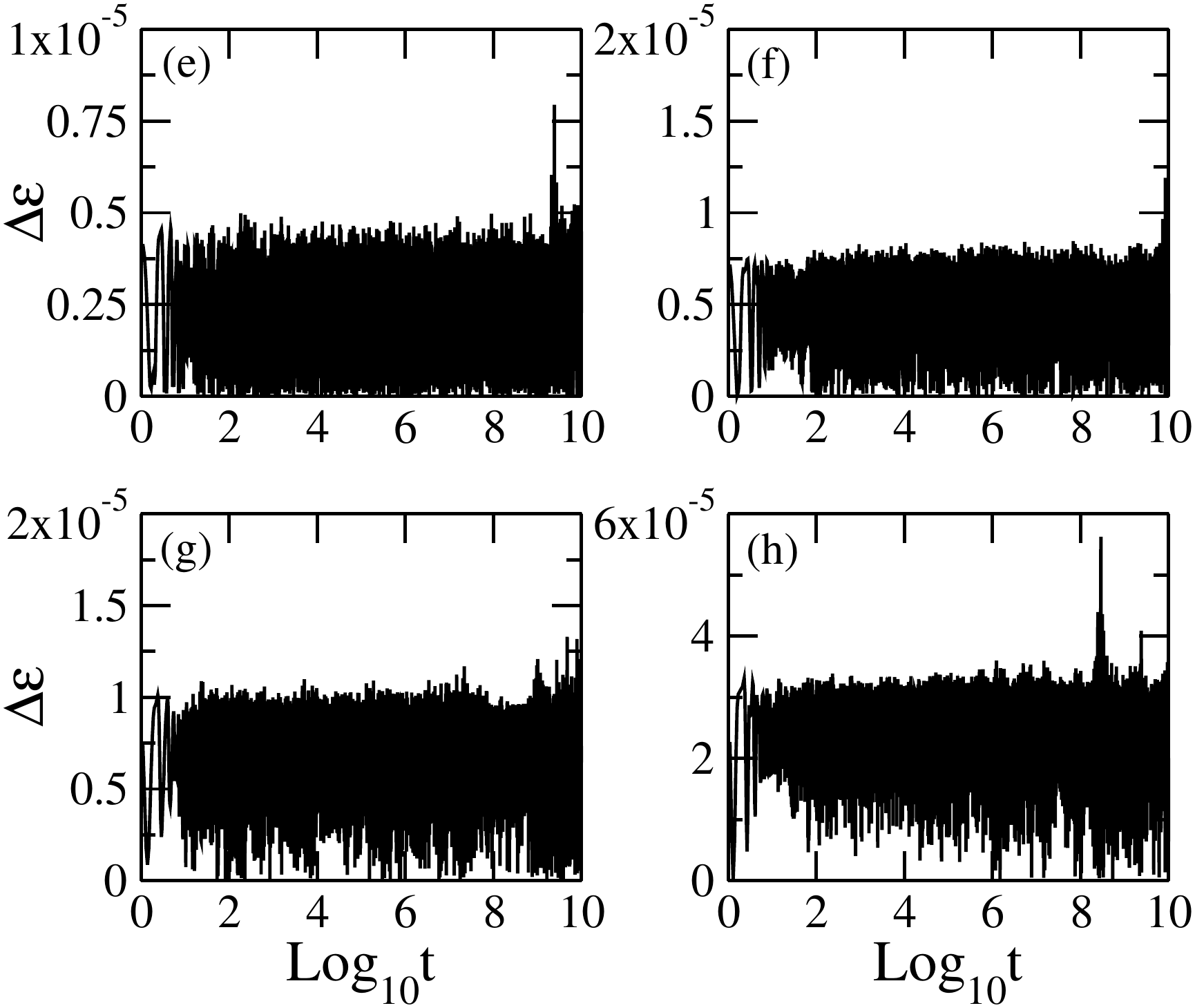}}
  \caption{Left plot: Time evolution of the relative energy error $\Delta \epsilon$ for the FPU chain, with
$\epsilon =  0.0091$ (a), 
$\epsilon =   0.0204$ (b),
$\epsilon =   0.0566$ (c) and 
$\epsilon =  0.145 $ (d) 
and time step $\tau = 0.1$.
Right plot: Time evolution of the relative energy error $\Delta \epsilon$ for the KG chain, with
$\epsilon =  1.371$ (e), 
$\epsilon =   1.629$ (f),
$\epsilon =   1.897$ (g) and 
$\epsilon =  3.501 $ (h) 
and time step $\tau = 0.1$.}
    \label{fig:energy_error_t0.1}
\end{figure}
In our simulations we used $\tau=0.1$ for both FPU and KG. With this time step, the $SABA_2C$ yields a relative error of the energy density $\Delta \epsilon = |\epsilon(t) - \epsilon(0)| /\epsilon(0) $ of the order of $10^{-5}$ (Fig.\ref{fig:energy_error_t0.1}). 
In KG case (right plot), we notice that even if it remains well below $10^{-5}$, the energy error $\Delta \epsilon$ shows an increase of the order of its fluctuations.

\section{Relation between the initial conditions and the energy density $\epsilon$} \label{sec:conv}

The initial condition of the FPU system chosen in our work corresponds to the FPU choice \cite{Fermi55}. It consists in exciting the lowest frequency mode in the space coordinate $q_n$ while keeping the conjugate momenta $p_n$ identically zero:
\begin{equation}
q_n(0) = A\sin\bigg(\frac{\pi n}{N+1}\bigg)\ ,\quad p_n(0)=0 \;.
\label{eq:FPU_IC}
\end{equation}
The total energy $E = H_\alpha(p_n(0),q_n(0))$ reads
\begin{equation}
E = N\frac{A^2\omega_1^2}{4} \ ,\quad \omega_1 = 2\sin\bigg(\frac{\pi }{2(N+1)}\bigg)\ .
\label{eq:total_energy}
\end{equation}
which yields the energy density $\epsilon$
\begin{equation}
\epsilon = \frac{E}{N} = \frac{A^2\omega_1^2}{4} \ ,
\label{eq:energy_density}
\end{equation}
In order to help the interested reader to compare data from different publications using different notations, we show
in Tab.\ref{tab:conversion_FPU} the values of the energy density $\epsilon$ we have considered in our work, and their corresponding values of the initial amplitude $A$:
\begin{table}[h!]
\centering
\begin{tabular}{
|P{1.7cm}|P{2.2cm}| }
\hline
Amplitude  $A$ &   Energy density $\epsilon$ \\
\hline 
1.0       &    0.0023\\
1.5       &     0.0051\\
1.6      &     0.0058\\
1.7   &     0.0065\\
1.8   &     0.0073\\
1.9      &     0.0082\\
2.0          &     0.0091\\
2.5      &         0.0142\\
3.0          &     0.0204\\
4.0        &      0.0362\\
5.0        &     0.0566\\
6.0        &       0.0815\\
7.0      &    0.111\\
8.0        &     0.145\\
9.0         &     0.183\\
\hline
\end{tabular} 
\caption{Conversion table between the values of the amplitude $A$  and the energy density $\epsilon$ for the FPU chain. }
\label{tab:conversion_FPU}
\end{table}

For the KG chain, we choose random initial conditions at all sites 
\begin{equation}
p_n(0)=c\cdot\xi_{2n-1}\ ,\quad q_n(0)=c\cdot\xi_{2n}\ , \qquad\xi_n\in[-0.5,0.5]\ ;
\label{eq:KG_IC}
\end{equation} 
where $\xi_n$ are uniformly distributed random values over the interval $[-0.5,0.5]$, and $c>0$. The energy density $\epsilon = E/N$ depends on the choice of the parameter $c$ and different disorder realizations. 

\section{Comparison of $T_{FPU}$ with the data from Ponno {\it et al.} \cite{Ponno11}} \label{sec:eps_vs_alpha}

In Ref.\cite{Ponno11}, the nonlinear parameter is $\alpha=0.33$. 
A simple rescaling relates the corresponding data to our results obtained for $\alpha=0.25$.
Consider $(p_n(t),q_n(t))$ to be a solution of the FPU chain for a given $\alpha$. Consider a different nonlinear parameter $\bar{\alpha}$. Given the ratio $\vartheta = \bar{\alpha} / \alpha$, we define a rescaling of the solutions
\begin{equation}
\bar{p}_n(t) = \frac{p_n(t)}{\vartheta}\ ;\quad \bar{q}_n(t) = \frac{q_n(t)}{\vartheta}\ .
\label{eq:rescaling}
\end{equation}
It follows that this trajectory solves the FPU Hamiltonian equations with $\bar{\alpha}$ as the nonlinear parameter. 
The amplitude of the initial state is rescaled as $\bar{A}=A/\theta$ and the energy (and the energy density) is
rescaled as $\bar{E} = E / \theta^2$.
\begin{figure}[h]
 \centering
 \includegraphics[ width=0.5\columnwidth]{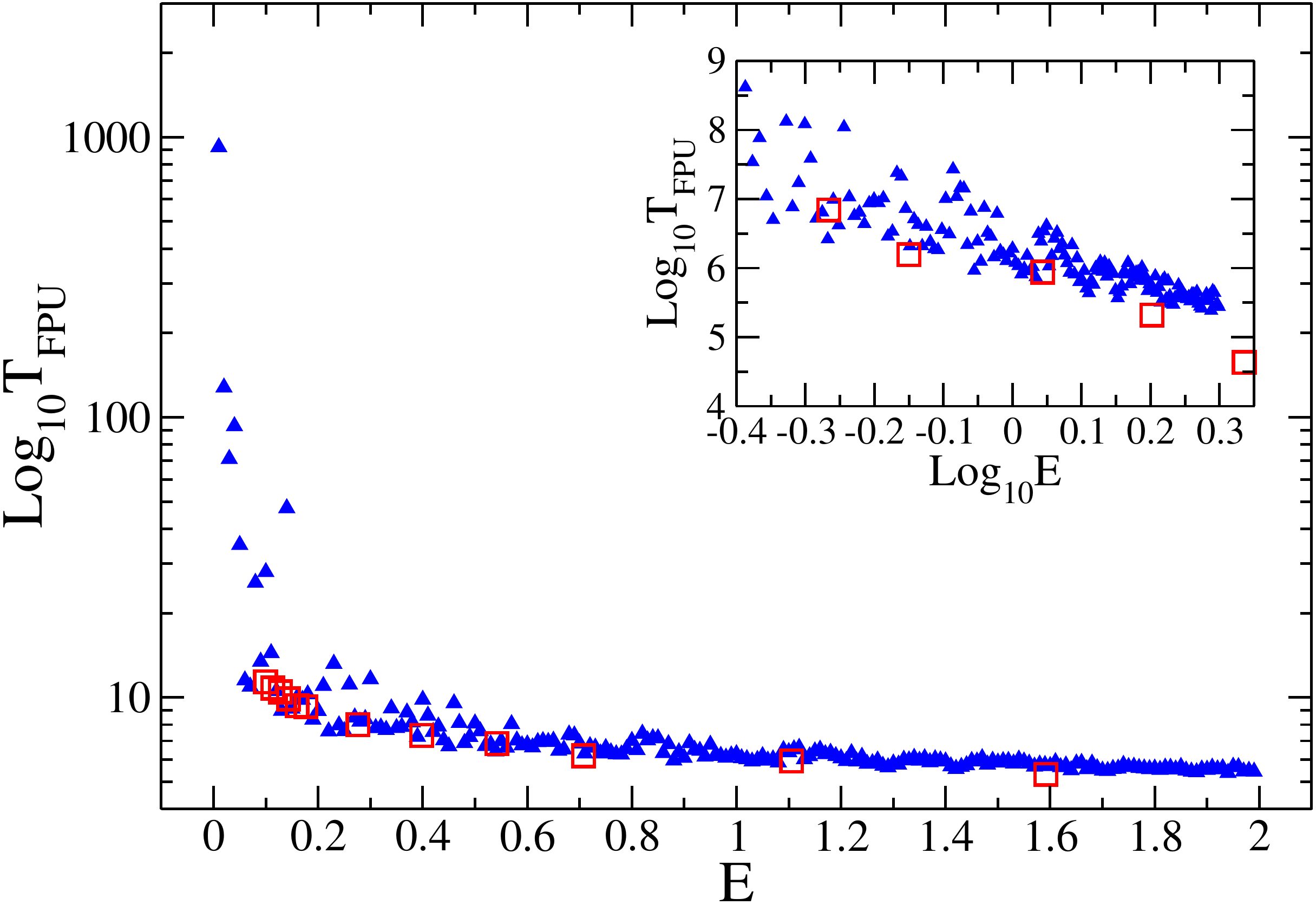}
 \caption{Equipartition time $Log_{10} T_{FPU}$ on a logarithmic scale vs. energy $E=N \epsilon$. Inset: equipartition time $Log_{10} T_{FPU}$ vs. $Log_{10} E$. Blue data courtesy of H. Christodoulidi (published in \cite{Ponno11}) obtained for $\alpha=0.33$. Red data computed by us for $\alpha=0.25$. System size $N=31$.}
 \label{eq:us and heleni}
\end{figure}
In Ref.\cite{Ponno11} a system size of $N=31$ was used; thus we recomputed our equipartition times $T_{FPU}$ for 
that case. The corresponding data in Ref.\cite{Ponno11} were obtained from an extrapolation of the energy flow
from strongly to weakly excited normal modes. 
In Fig.\ref{eq:us and heleni} we compare  our equipartition times $T_{FPU}$ for $\alpha=0.25$ with those from Ponno {\it et.al.} for $\alpha=0.33$.  
The blue triangles (Ponno {\it et.al.} data) and the red squares (our data) are in very good agreement.

\section{CPU type}\label{sec:CPU}

Numerical simulations presented in the paper has been performed on the Massey Cluster Simurg, which uses Intel Xeon CPU E5-2670 processors; and on the PCS-IBS Cluster, which uses Intel E5-2680v3 processors.

\section{Stickiness to regular orbits} \label{sec:stickiness}

\begin{figure}[h]
 \centering
 \includegraphics[ width=0.5\columnwidth]{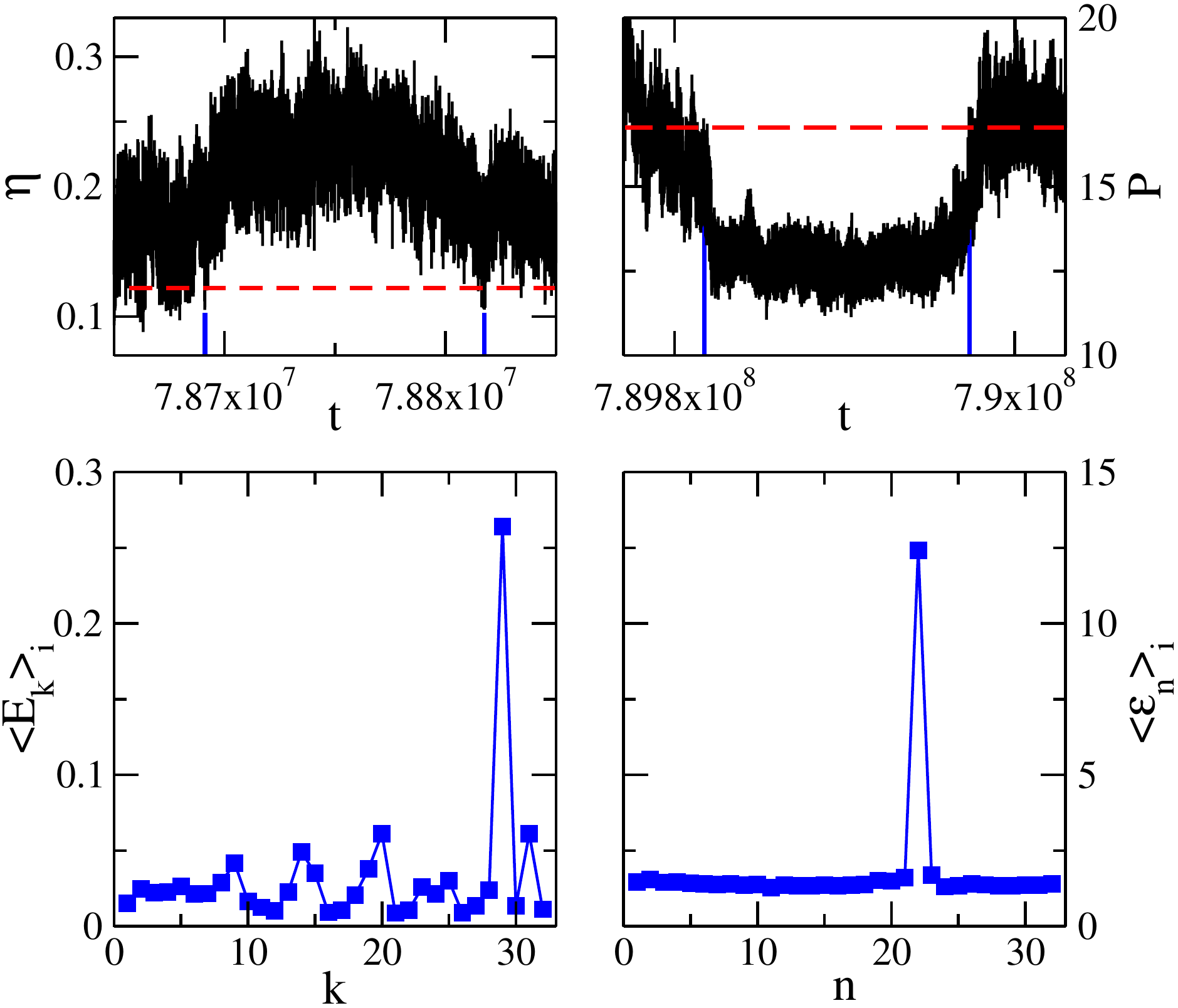}
 \caption{
Upper plots: Time evolution of the observables $\eta$ (left) and $P$ (right) of the trapping events shown in Fig.3 of the main text for FPU (left) and KG (right). The red dashed lines correspond to the equipartition values of the entropy $\langle\eta\rangle=0.1218$ and $\langle P \rangle=16.7473$, and the blue ticks mark the consecutive crossings of the manifolds $\mathcal{F}_{\eta}$ and $\mathcal{F}_{P}$ respectively.
Bottom plots: The mode energies $\langle E_k \rangle_i$ (left) and energy densities $\langle \epsilon_n \rangle_i $ (right) averaged over the two events in the upper plots versus mode number $k$ and lattice site $n$ respectively.
}
 \label{fig:orbits}
\end{figure}

%
In the upper row of Fig.\ref{fig:orbits} we show the time evolution of the observables $\eta$ and $P$ in correspondence of the excursion far from equilibrium shown in Fig.3 of the main text for the FPU (left plots) and the KG (right plots).
At the beginning and the end of the return times marked by the blue ticks in Fig.3 of the main text, the formation and the dissolution of the inhomogeneous distributions takes place.
In the lower plot of Fig.\ref{fig:orbits} we plot the time average of the energies over the return times 
\begin{equation}
\begin{split}
\langle E_k \rangle_i &= \frac{1}{t_r(i)}\int_{t_i}^{t_{i+1}} E_k(t)dt\ ,\quad k=1,\dots, N\\
\langle \epsilon_n \rangle_i  &= \frac{1}{t_r(i)}\int_{t_i}^{t_{i+1}} \epsilon_n(t)dt\ , \quad  n=1,\dots, N\ .
\end{split}
\label{recaverage}
\end{equation}
These averages highlight that the inhomogeneous excitations shown in Fig.3 of the main text are not sparse peaks, which otherwise would be erased on time average, but instead are consistent breather-like excitations that persist for the whole time of the excursion. 

\section{Fitting $\delta$ in the tails of $\mathcal{P}_+(t_r)$} \label{sec:PDF}

We obtain the PDF using bins equispaced on a logarithmic scale. This choice allows a more precise evaluation of the exponent $\delta$. Note that the conversion to the correct equidistant binning on a linear scale changes
the exponent by 1.
The correct function is obtained by multiplying the log-binned PDF by $1 / t_r$.  We estimate the exponent $\delta$ by a power-law regression of the function $\mathcal{P}_\pm({t_r})$ performed on intervals of decreasing length $I$ in the PDF tail, by fixing the upper interval edge $t_r^{\text{MAX}}$ 
and varying the lower edge $t_r^{\text{MIN}}$.

\begin{figure}[h!]
 \centering
  \includegraphics[ width=0.47\columnwidth]{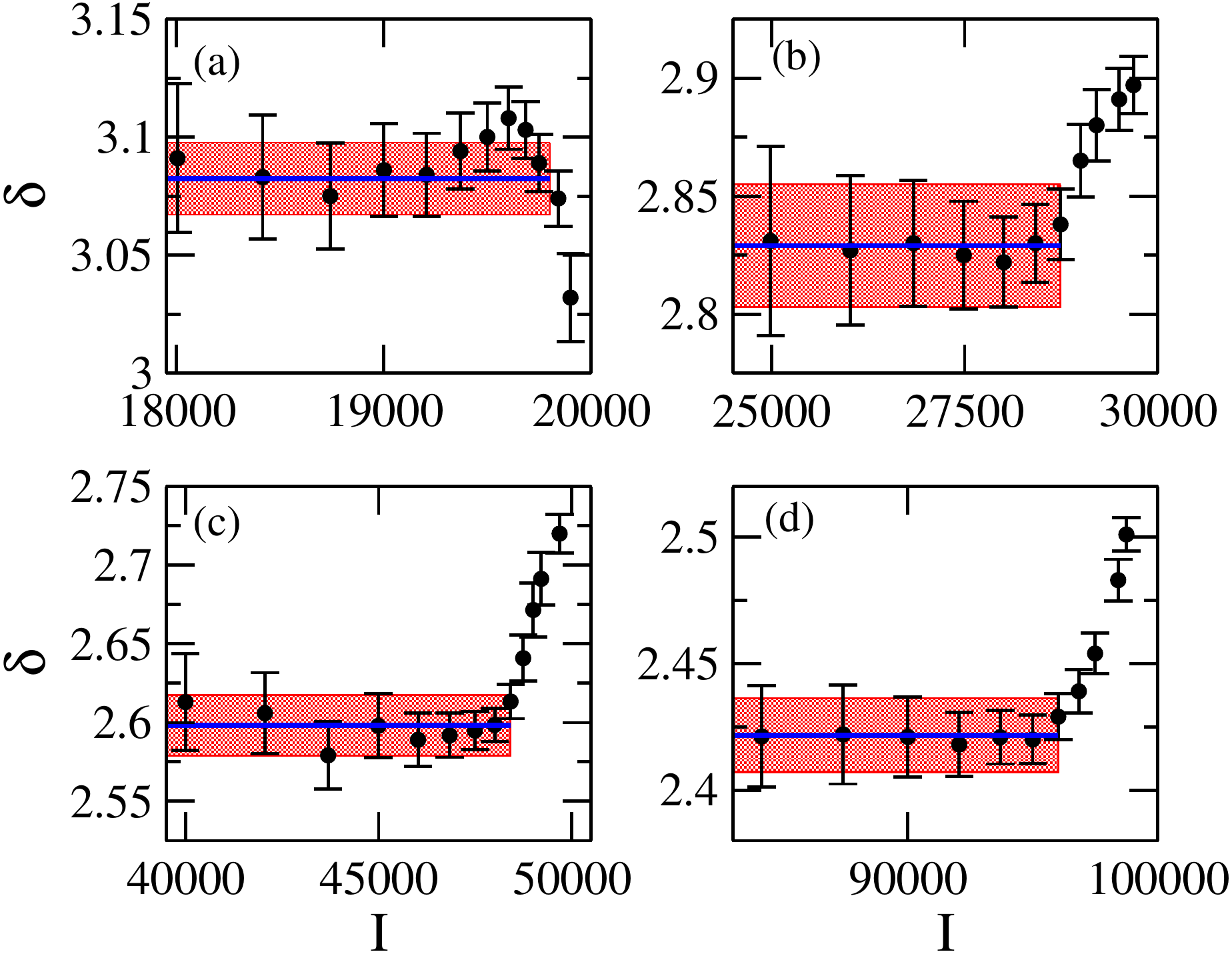}
  \includegraphics[ width=0.46\columnwidth]{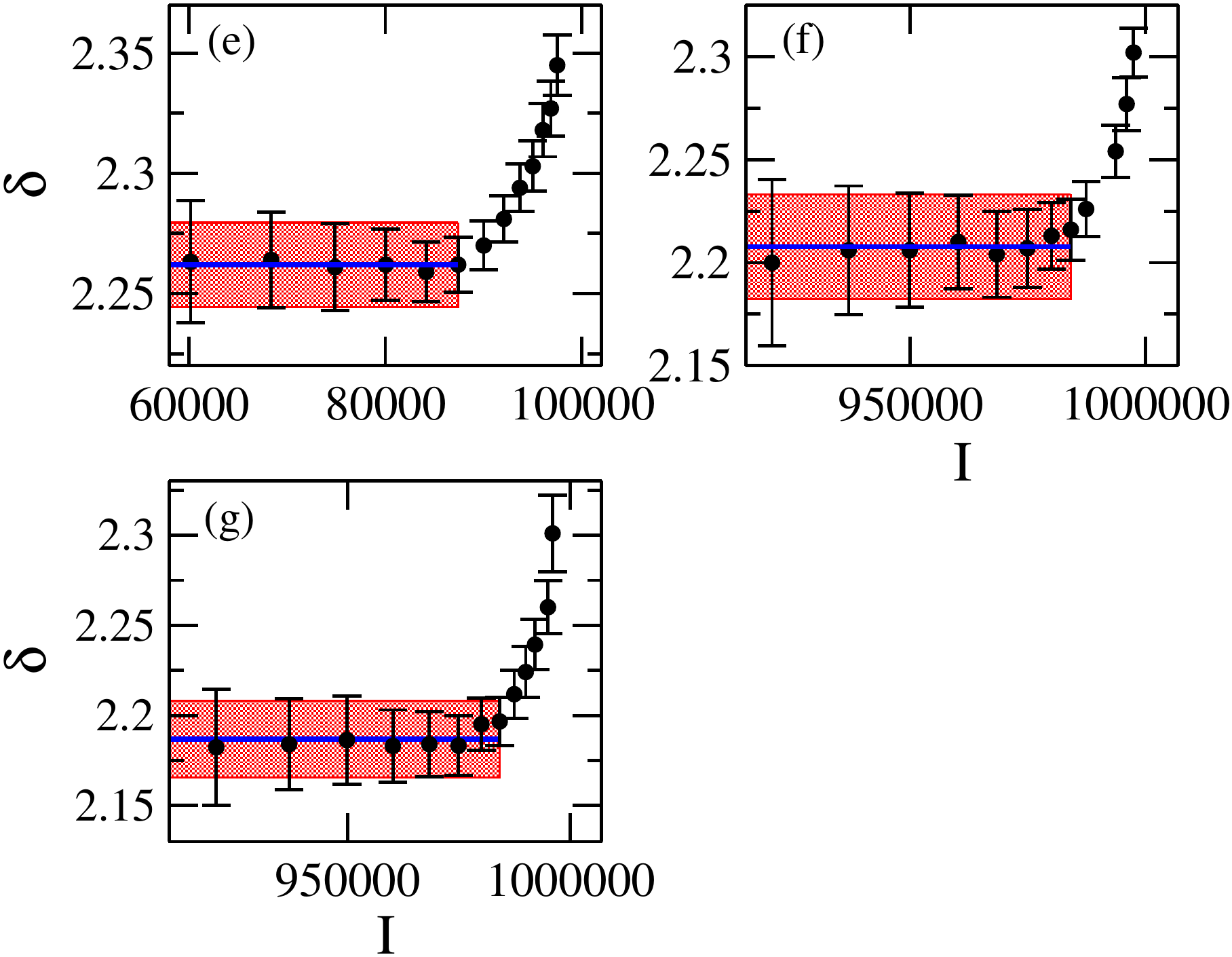}
  \caption{Estimating the exponent $\delta$ for
$\epsilon = 0.145$ (a),
$\epsilon = 0.111$ (b),
$\epsilon = 0.0815$ (c),
$\epsilon = 0.0566$ (d),
$\epsilon = 0.0362$ (e),
$\epsilon = 0.0204$ (f) and
$\epsilon = 0.0091$ (g).   
The blue lines indicate the average values of the exponent $\delta$, while the red shaded areas denote the standard deviations.} 
  \label{fig:exponents_vs_eps_FPU}
\end{figure}

In Fig.\ref{fig:exponents_vs_eps_FPU} we show the computations of the estimates for the FPU chain for energy densities $\epsilon$ ranging from $0.0204$ to $0.145$.

\begin{figure}[h!]
 \centering
  \includegraphics[ width=0.47\columnwidth]{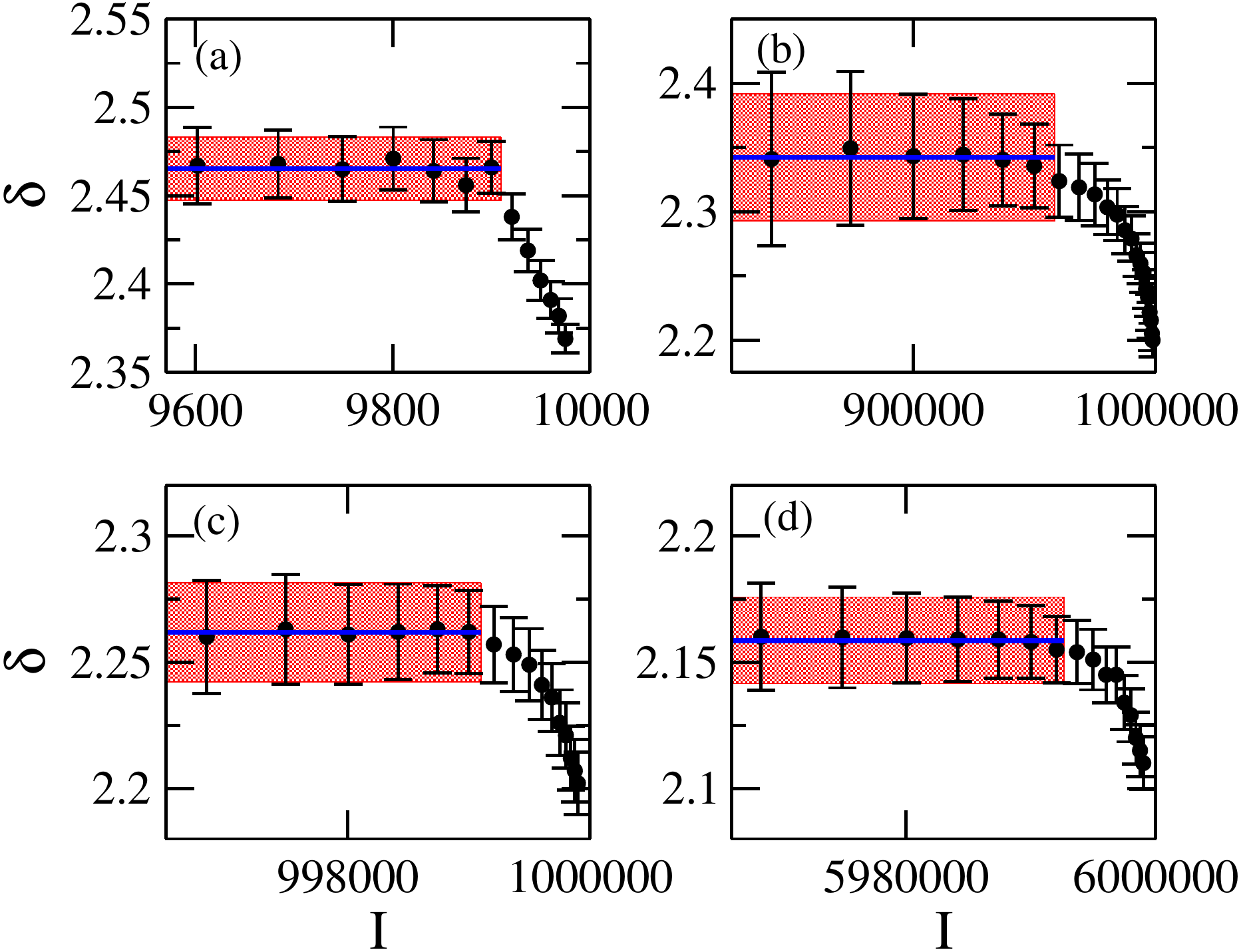}
  \includegraphics[ width=0.46\columnwidth]{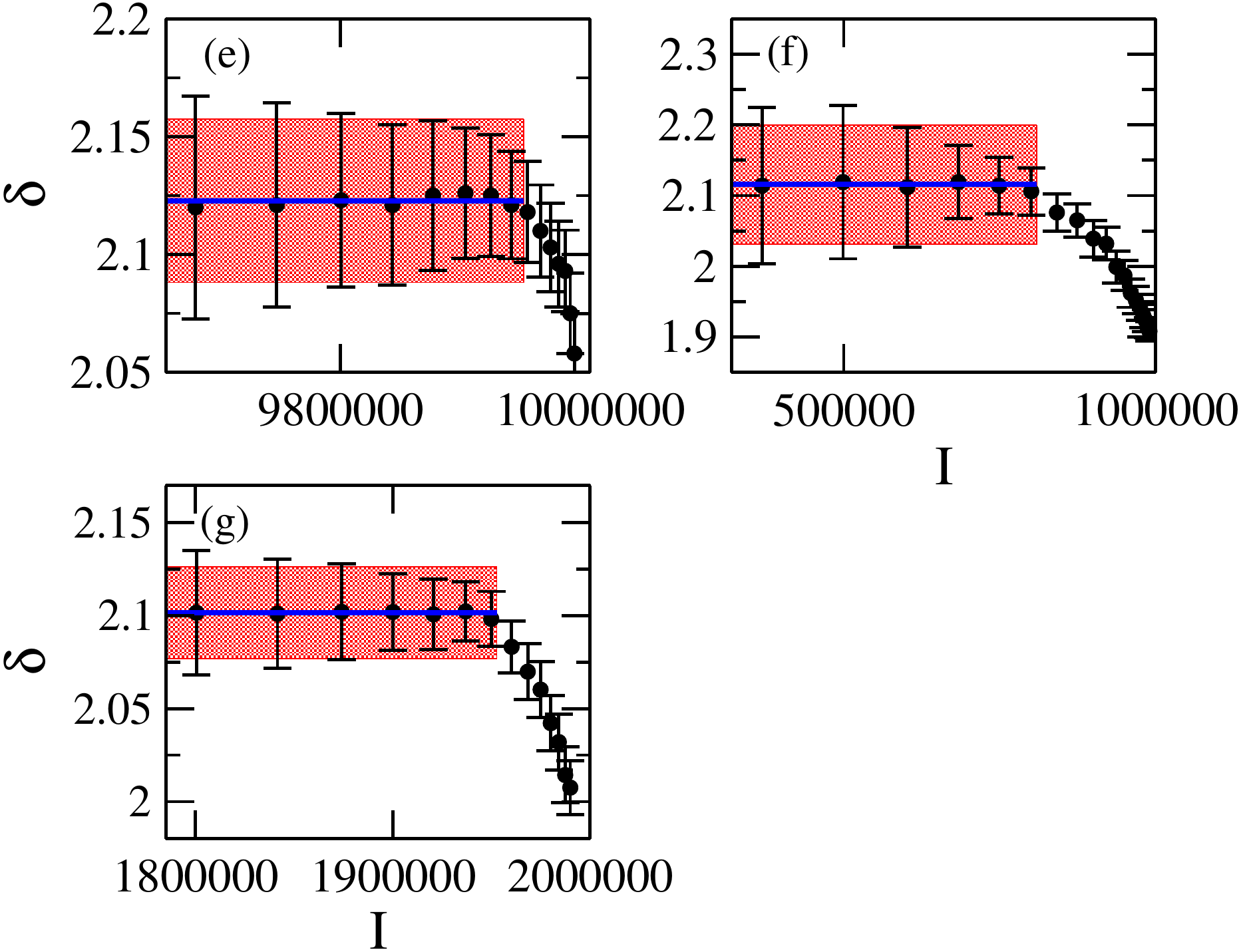}
  \caption{Estimating the exponent $\delta$ for
$\epsilon = 1.184 $ (a),
$\epsilon = 1.303$ (b),
$\epsilon = 1.416$ (c),
$\epsilon = 1.867$ (d),
$\epsilon = 2.149$ (e),
$\epsilon = 2.655 $ (f) and
$\epsilon = 3.166$ (g).   
The blue lines indicate the average values of the exponent $\delta$, while the red shaded areas denote the standard deviations.} 
  \label{fig:exponents_vs_eps_KG}
\end{figure}

In Fig.\ref{fig:exponents_vs_eps_KG} we show the computations of the estimates for the KG chain for energy densities $\epsilon$ ranging from $1.184$ to $3.166$.

\end{document}